\documentclass[aps, prd, amsmath, floats, floatfix, twocolumn, nofootinbib, superscriptaddress, showpacs]{revtex4-1}
\usepackage{graphicx}
\usepackage{color}
\usepackage{latexsym}

\newcommand{\be}{\begin{eqnarray}}
\newcommand{\ee}{\end{eqnarray}}

\begin{document}

\title{Testing the Kerr black hole hypothesis:\\
comparison between the gravitational wave and the iron line approaches}

\author{Alejandro~C\'ardenas-Avenda\~no}
\affiliation{Center for Field Theory and Particle Physics and Department of Physics, Fudan University, 200433 Shanghai, China}
\affiliation{Programa de Matem\'atica, Fundaci\'on Universitaria Konrad Lorenz, 110231 Bogot\'a, Colombia}

\author{Jiachen Jiang}
\affiliation{Center for Field Theory and Particle Physics and Department of Physics, Fudan University, 200433 Shanghai, China}

\author{Cosimo Bambi}
\email[Corresponding author: ]{bambi@fudan.edu.cn}
\affiliation{Center for Field Theory and Particle Physics and Department of Physics, Fudan University, 200433 Shanghai, China}
\affiliation{Theoretical Astrophysics, Eberhard-Karls Universit\"at T\"ubingen, 72076 T\"ubingen, Germany}

\date{\today}

\begin{abstract}
The recent announcement of the detection of gravitational waves by the LIGO/Virgo collaboration has opened a new window to test the nature of astrophysical black holes. Konoplya \& Zhidenko have shown how the LIGO data of GW~150914 can constrain possible deviations from the Kerr metric. In this letter, we compare their constraints with those that can be obtained from accreting black holes by fitting their X-ray reflection spectrum, the so-called iron line method. We simulate observations with eXTP, a next generation X-ray mission, finding constraints much stronger than those obtained by Konoplya \& Zhidenko. Our results can at least show that, contrary to what is quite commonly believed, it is not obvious that gravitational waves are the most powerful approach to test strong gravity. In the presence of high quality data and with the systematics under control, the iron line method may provide competitive constraints.
\end{abstract}


\maketitle


\section{Introduction}

In the past 50~years, Solar System experiments and radio observations of binary pulsars have tested general relativity in the weak field limit~\cite{will}. The focus of current experiments has now shifted to testing the theory in other regimes. In this context, an important line of research is devoted to confirm the nature of astrophysical black holes~\cite{rev,rev-iron,rev-tj1,yagi,rev-tj2}. In 4-dimensional general relativity, an uncharged black hole is described by the Kerr solution, which is completely characterised by the mass $M$ and the spin angular momentum $J$ of the compact object. The spacetime geometry around astrophysical black holes is expected to be very close to the Kerr metric and therefore possible deviations from the Kerr solution could only arise from new physics~\cite{rev}.

The LIGO/Vigo collaboration has recently announced the detection of gravitational waves from the coalescence of two stellar-mass black holes~\cite{ligo}. The event was named GW~150914. The mass of the final black hole is estimated to be $M = 62^{+4}_{-4}$~$M_\odot$ and its dimensionless spin parameter would be $a_* = 0.67^{+0.05}_{-0.07}$. The possibility of detecting gravitational waves opens now a new window to study general relativity in the strong field regime and to test the nature of astrophysical black holes~\cite{rgw,ligo-gr}.

In Ref.~\cite{kz}, Konoplya \& Zhidenko have employed a parametrised approach to get a simple estimate on how the data of GW~150914 can test the nature of black holes and constrain possible deviations from the Kerr solution. Their parametrisation is based on the proposal of Ref.~\cite{krz}. The line element reads
\be\label{eq-m}
ds^2 &=& - \frac{N^2(r,\theta) - W^2(r,\theta) \sin^2\theta}{K^2(r,\theta)} dt^2
\nonumber\\ &&
- 2 W(r,\theta) \, r \sin^2\theta \, dt d\phi
+ K^2(r,\theta) \, r^2 \sin^2\theta \, d\phi^2
\nonumber\\ &&
+ \Sigma(r,\theta) \left[\frac{B^2(r,\theta)}{N^2(r,\theta)}dr^2 + r^2d\theta^2\right] \, ,
\ee
where
\be\label{eq-m2}
N^2(r,\theta) &=& \frac{r^2 - 2Mr + a^2}{r^2} - \frac{\eta}{r^3} \, , \nonumber\\
B^2(r,\theta) &=& 1 \, , \nonumber\\
\Sigma(r,\theta) &=& \frac{r^2 + a^2 \cos^2\theta}{r^2} \, , \nonumber\\
K^2(r,\theta) &=& \frac{\left(r^2 + a^2\right)^2 - a^2 \sin^2\theta 
\left(r^2 - 2Mr + a^2\right)}{r^2 \left(r^2 + a^2 \cos^2\theta\right)}
\nonumber\\ &&
+ \frac{\eta a^2 \sin^2\theta}{r^3 \left(r^2 + a^2 \cos^2\theta\right)}\, , \nonumber\\
W(r,\theta) &=& \frac{2Ma}{r^2 + a^2 \cos^2\theta} 
+ \frac{\eta a}{r^2 \left(r^2 + a^2 \cos^2\theta\right)} \, ,
\ee
and $a = J/M$ is the rotation parameter. The deformation parameter $\eta$ can be written as
\be
\eta = r_0 \left(r_0^2 - 2 M r_0 + a^2 \right) \, ,
\ee
where $r_0$ is the radial coordinate of the event horizon of the black hole metric in~(\ref{eq-m}). If we write
\be
r_0 = r_{\rm Kerr} + \delta r = M + \sqrt{M^2 - a^2} + \delta r \, ,
\ee
we can use $\delta r$ as the deformation parameter to quantify possible deviations from the Kerr spacetime. If $\delta r = 0$, $r_0$ reduces to the radial position of the event horizon of a Kerr black hole. In the general case, $\delta r$ measures the difference of the radial coordinate of the event horizon with respect to that of a Kerr black hole with the same mass and spin.

The choice of the metric in Eq.~(\ref{eq-m}) is somewhat arbitrary, but it posses some nice properties. It significantly simplifies the calculations of the quasi-normal modes, it is close to 
the Kerr metric at large distances (it has the post-Newtonian parameters $\beta = \gamma = 1$ as the Schwarzschild solution and the same mass quadrupole moment as the Kerr spacetime $Q = - a^2 M$), and it is quite different near the horizon~\citep{kz}.

In the case of the coalescence of two black holes in a binary, we can distinguish three stages: the post-Newtonian inspiral, the merger, and the ringdown. In their analysis, Konoplya \& Zhidenko only study the ringdown phase, arguing that it is the most suitable to test strong gravity. The post-Newtonian inspiral can indeed be the same in many alternative theories of gravity, and their metric is as the Kerr metric in the asymptotic region. The merger is a very short and complicated stage. As a further simplification, they study the quasi-normal frequencies of a scalar field in the deformed background. In general relativity and in other theories of gravity, these frequencies are not much different from those of the gravitational waves derived from the field equations. However, this is not always true and there are also examples of gravity theories in which the scalar field quasi-normal frequencies can be quite different from those of the gravitational waves. With such a simplification, the quasi-normal frequencies $\omega$ only depend on the background metric and on the quantum numbers $l$ and $m$; that is~\citep{kz}
\be
\omega^2 = \omega^2 (M, a, \delta r, m, l) \, .
\ee

Within the WKB approximation~\citep{sw,k}, the $m=l=2$ mode scalar field quasi-normal frequency in a Kerr background with $a_* = 0.65$ is
\be
\omega M = 0.635 - 0.0901 \, i \, .
\ee 
Fig.~1 in Ref.~\cite{kz} shows that there is a degeneracy between the spin $a_*$ and possible deviations from the Kerr metric $\delta r$. In other words, the same quasi-normal frequency may be associated either to a Kerr black hole with spin $a_* = 0.65$ or to a non-Kerr black hole with a different spin. Under this scope, the seminal detection of GW~150914 does not seem so strong and large deviations from the Kerr solution, at the level of $\delta r/r_{\rm Kerr} \approx -0.8$, cannot be ruled out.

Generally speaking, the frequency of a quasi-normal mode is characterised by a real and an imaginary part, namely two parameters. If we measure only one frequency (as in the case of GW~150914) and we assume the Kerr metric, this is enough to estimate $M$ and $a_*$. If we relax the Kerr black hole hypothesis, we have at least one deformation parameter, and there may be a degeneracy. In the case of gravitational waves, the degeneracy may be broken with the measurement of other modes, but it depends on the metric. For instance, the measurement of other modes does not seem to be very helpful to distinguish Kerr and Kerr-Sen black holes~\cite{kz}.

\begin{figure*}[t]
\vspace{0.5cm}
\begin{center}
\includegraphics[type=pdf,ext=.pdf,read=.pdf,width=7.5cm]{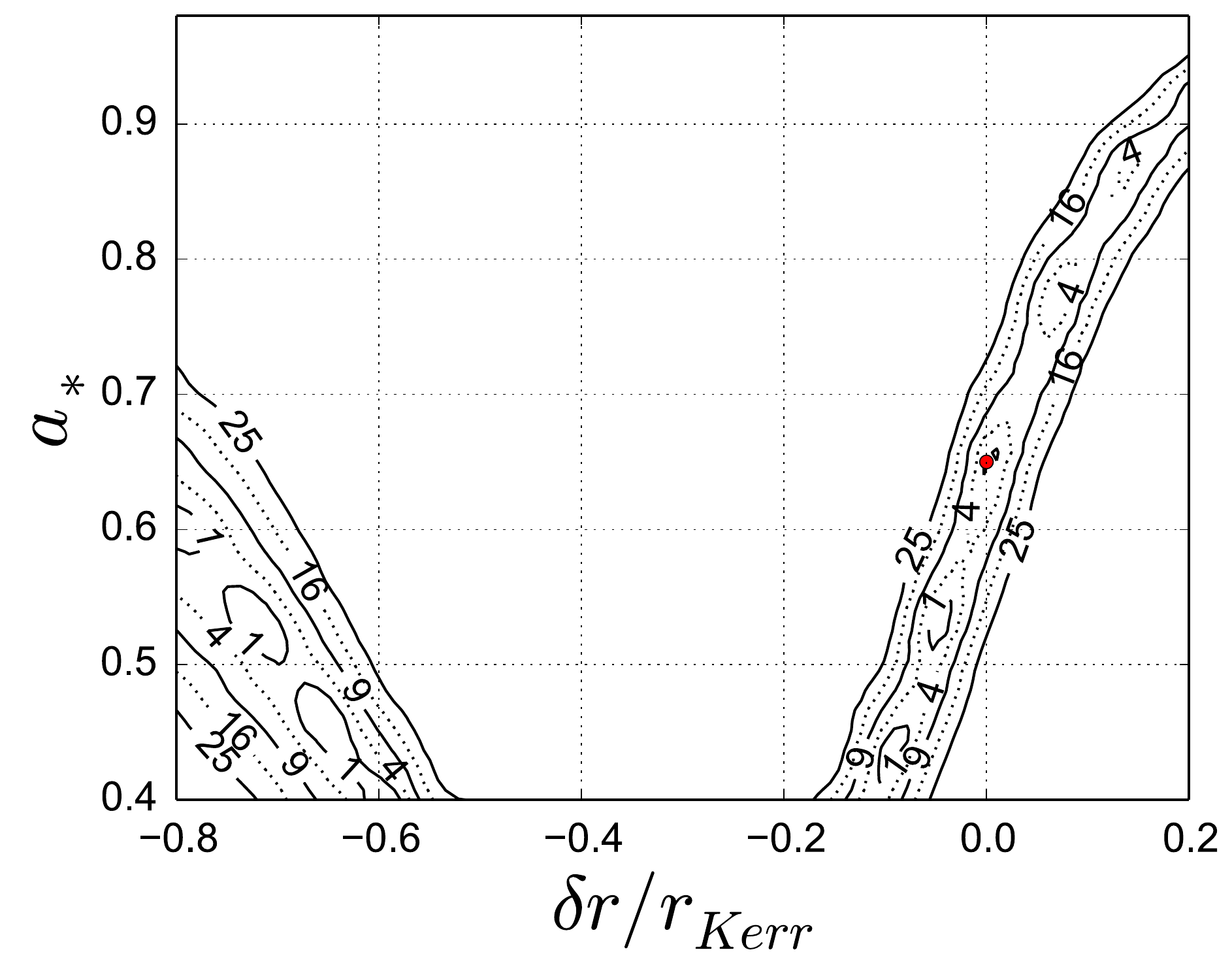}
\hspace{0.5cm}
\includegraphics[type=pdf,ext=.pdf,read=.pdf,width=7.5cm]{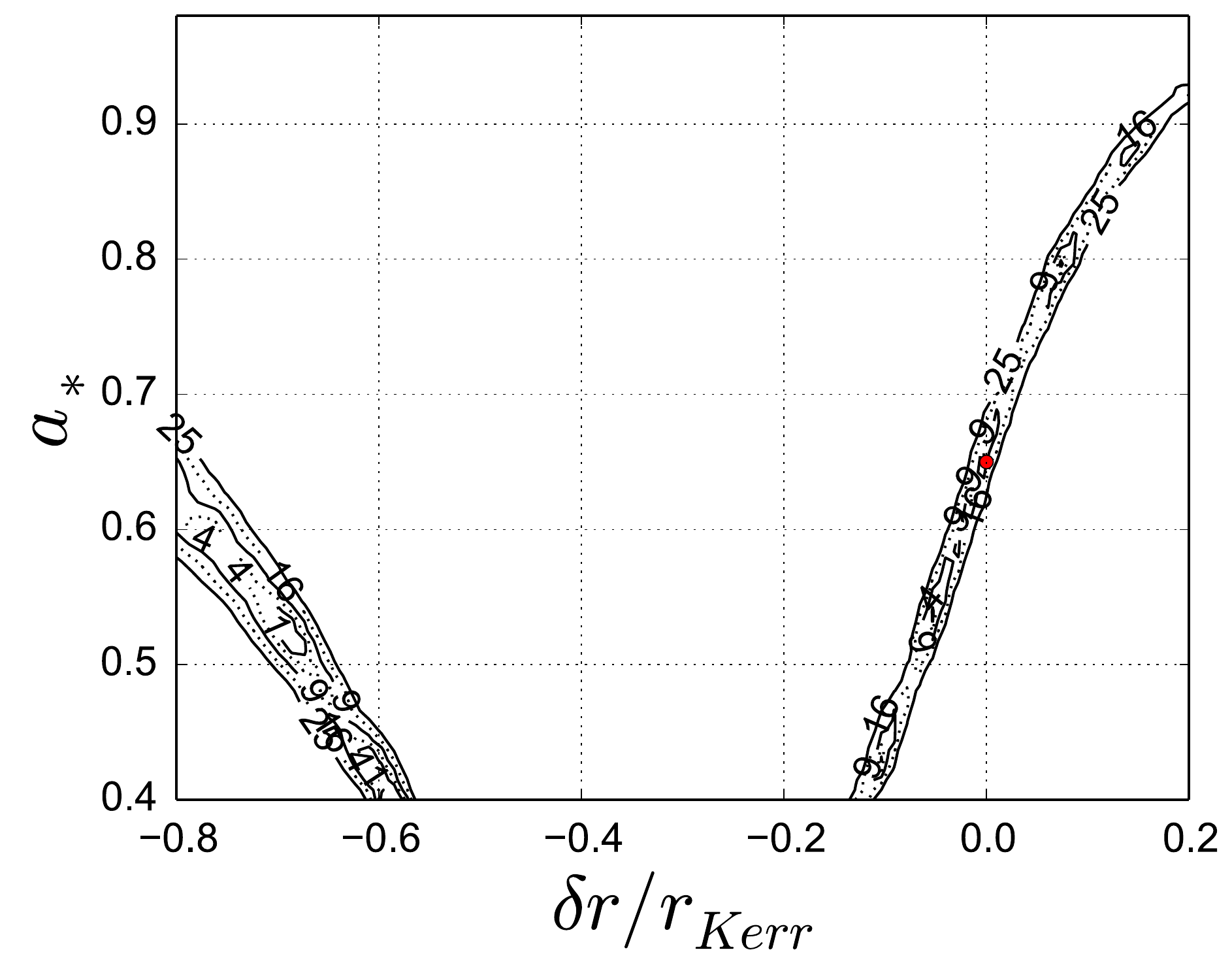}
\end{center}
\vspace{-0.4cm}
\caption{$\Delta\chi^2$ contours with $N_{\rm line} = 10^4$ (left panel) and $10^5$ (right panel) from the comparison of the iron line profile of a Kerr black hole simulated using an input spin parameter $a_*' = 0.65$ and an inclination angle $i' = 45^\circ$ vs a set of non-Kerr black holes with spin parameter $a_*$ and deformation parameter $\delta r/r_{\rm Kerr}$. The red dot indicates the reference black hole. See the text for more details. \label{fig1}}
\end{figure*}

\begin{figure}[h]
\begin{center}
\includegraphics[type=pdf,ext=.pdf,read=.pdf,width=7.5cm]{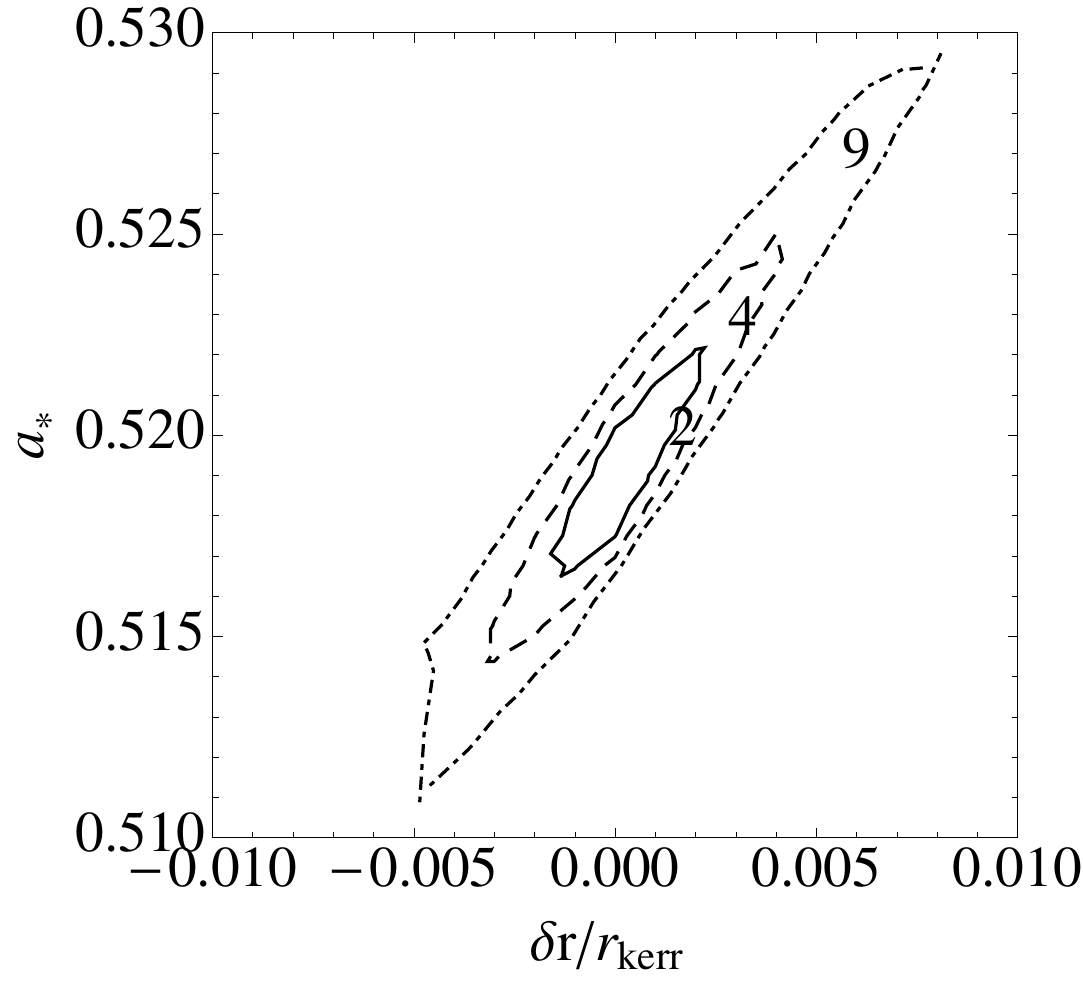}
\end{center}
\vspace{-0.4cm}
\caption{$\Delta\chi^2$ contours from the simulation of a 42~ks observation of XTE~J1752-223 with eXTP assuming that the source is a Kerr black hole with the parameters found in Ref.~\cite{xtej1752}. See the text for more details. \label{fig-extp}}
\end{figure}

\begin{figure*}[t]
\vspace{0.5cm}
\begin{center}
\includegraphics[type=pdf,ext=.pdf,read=.pdf,width=16cm]{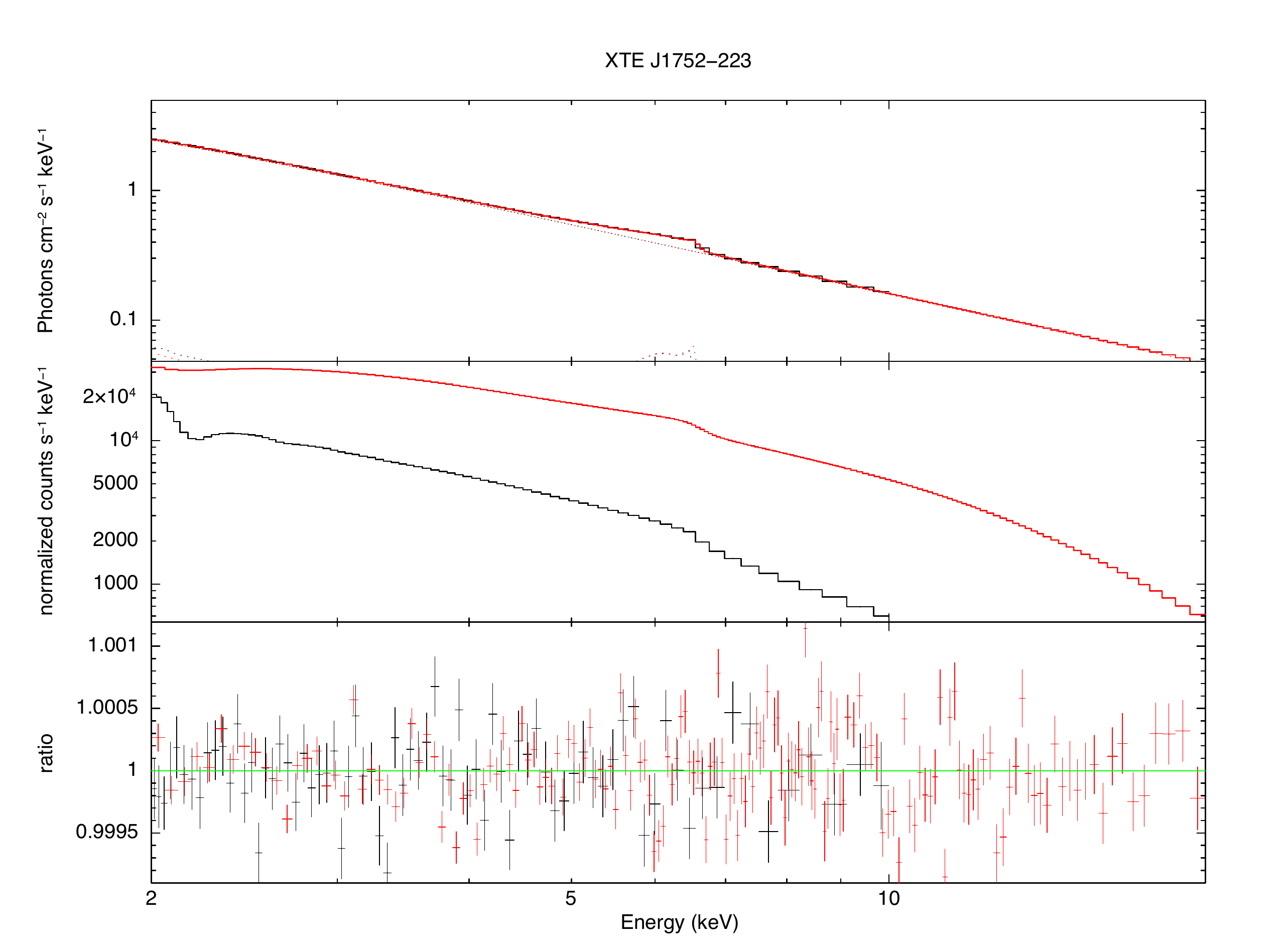}
\end{center}
\vspace{-0.4cm}
\caption{Spectrum of XTE~J1752-223. Top panel: simulated unfolded spectrum of an XTE~J1752-223-like source. Middle panel: simulated LFA and LAD spectra of XTE~J1752-223 for a 42~ks exposure based on the spectral model for XTE~J1752-223. Bottom panel: ratio to the same model combination. The black data refer to the simulated observation of LFA and the red data to the ones of LAD. The solid lines are for the total flux. The dotted lines indicate the power-law component, the reflection component, and a soft thermal component of the disk which is mainly below 2~keV and is ignored. See the text for more details. \label{fig-spectrum}}
\end{figure*}

Here we want to compare these constraints from gravitational waves from those that can be obtained from X-ray data of accreting black holes. In particular, we are interested in constraints from the iron line method~\cite{rey14}. Broad iron K$\alpha$ lines are commonly observed in the X-ray spectrum of black holes of all masses. They are generated by illumination of a cold accretion disk by a hot corona. The analysis of the iron line actually involves the fitting of the whole reflected component of the disk, but eventually the feature that can probe the spacetime geometry in the strong gravity region is mainly the iron K$\alpha$ line. Studies to test the Kerr metric with the iron line have been reported in~\cite{code2,cfmiron,jjc1,jjc2,acajc1,jp}, where it is shown that this technique is potentially more powerful than the analysis of the thermal spectrum to test black holes~\cite{cfm,code1,lingyao}. For a review on the use of the two techniques to test the Kerr metric, see e.g. Ref.~\cite{rev-iron}.

\section{X-ray reflection spectroscopy}

\subsection{Iron line method}

With the same spirit as the authors of Ref.~\cite{kz}, here we perform a simplified analysis to compare the constraining capability of gravitational waves and iron line to test the nature of black holes. Since we want to compare our results with those in Ref.~\cite{kz}, we use the metric in Eq.~(\ref{eq-m}) and the deformation parameter $\delta r$. Roughly speaking, we can expect that current iron line measurements are at the same level as the LIGO data of GW~150914, because the spin parameter measurement of the final black hole in GW~150914 has an uncertainty comparable to that obtainable from iron line measurements. However, the two techniques are sensitive to quite different relativistic effects.

As a very preliminary study, we follow the approach of Ref.~\cite{jjc2}. We compute the photon flux of the iron line profile in different background metrics over a grid of energy bins with resolution $\Delta E = 100$~eV in the range 1-9~keV. For simplicity, we fix the emissivity index $q$, for both the simulations and the model fitting, to 3, which corresponds to the Newtonian limit for lamppost coronal geometry. The iron line profile is added to a power-law continuum, which is normalised to include 100~times the number of iron line photons when integrated over the energy range 1-9~keV. The photon index of the continuum is $\Gamma'=2$. Depending on the line shape, this corresponds to a line equivalent width of $EW \approx 370$-440~eV.

Our simulated spectra include Poisson noise, are then binned to achieve a threshold of counts $n = 20$, and fitted employing $\chi^2$ analysis. Fig.~\ref{fig1} shows the results of these simulations. The reference model is a Kerr black hole with spin parameter $a_*' = 0.65$, as in the study in~\cite{kz}, and observed from a viewing angle $i' = 45^\circ$. We assume that the number of photon counts in the iron line photon is $N_{\rm line} = 10^4$ (left panel) and $N_{\rm line} = 10^5$ (right panel). In these fits, we have five free parameters: the spin parameter $a_*$, the deformation parameter $\delta r$, the viewing angle $i$, the ratio between the continuum and the iron line photon flux $K$, and the index of the continuum $\Gamma$.

These plots can be compared with the plot presented in Ref.~\cite{kz}. Both our analysis and that in 
Ref.~\cite{kz} are based on a simple model, so the constraints should be taken as a general guide. Nevertheless, it is remarkable that the constraining capability seems to be comparable. In our case, $EW \approx 400$~eV is quite an optimistic observation. For a bright AGN, current observations may have $N_{\rm line} \approx 10^3$, but in the case of a bright black hole binary we may already have $N_{\rm line} \approx 10^5$ or more.

\subsection{Simulation of an observation with eXTP}

To do a step further, we constrain the deformation parameter $\delta r/r_{\rm Kerr}$ by simulating an observation with eXTP, which is a China-Europe proposal for a next generation X-ray mission~\cite{extp}. For this purpose, we have chosen the stellar-mass black hole XTE~J1752-223. The spin parameter of this object was measured via the iron line method in Ref.~\cite{xtej1752} and was found to be $a_* = 0.52 \pm 11$\footnote{Among the spin measurements reported in the literature of stellar-mass black holes, there is no source with a spin estimate very close to $a_* = 0.65$, see e.g. Tab.~1 in Ref.~\cite{rev-iron}. The spin measurement of XTE~J1752-223 is not too far.}.

The 2009 outburst of XTE~J1752-223 was observed by XMM-Newton and Suzaku. Its flux in the energy range 2-10~keV was $\sim 1.1 \cdot 10^{-8}$~erg~s$^{-1}$~cm$^{-2}$~\cite{xtej1752}. Our simulations are based on the parameters reported in~\cite{xtej1752}. In particular, we use their equivalent width of $EW \approx 221$~eV. While eXTP spectra are simulated with same source brightness and same exposure time (42~ks), they are able to reach a much higher total photon count, about $8 \cdot 10^9$ during the whole exposure. The photon count in the iron line is $N_{\rm line} \sim 10^7$, and this explains why we obtain much better constraints than those in Fig.~\ref{fig1} (see below).

The results of these simulations are shown in Figs.~\ref{fig-extp} and \ref{fig-spectrum}. We have used Xspec with the model~\cite{xillver1,xillver2} 
\begin{quote}
\verb|TBabs*(powerlaw+relconv*xillver+NKL)|
\end{quote}
where \verb|NKL| is the iron line in the non-Kerr metric in~(\ref{eq-m2}) from our code~\cite{code1,code2}. \verb|xillver| already includes the iron line, but we have kept its normalisation low to have a reasonable equivalent width in the iron line produced by our code for the non-Kerr metric. We have also tried the models
\begin{quote}
\verb|TBabs*(powerlaw+NKL)|
\verb|TBabs*(powerlaw+relconv*pexrav+NKL)|
\end{quote}
without finding substantial differences. The model without reflection component and only our \verb|NKL| clearly provides stronger constraints, which are likely too optimistic. In the other two cases (\verb|xillver| and \verb|pexrav|), without a correct non-Kerr reflected component, the final constraint may be slightly worse than what we can actually obtain, as other features should also change in a non-Kerr background and provide some help to constrain the spacetime geometry. We have not added any thermal component to our model because the range of our spectrum is 2-10~keV (LFA) and 2-20~keV (LAD). The spectra are rebinned to have a minimum count of 20 photons per bin. The plots in Fig.~\ref{fig-spectrum} are also rebinned to have nicer pictures.

The constraints in Fig.~\ref{fig-extp} look quite strong and, unlike the constraints in Fig.~1 in~\cite{kz} from GW~150914 and those in Fig.~\ref{fig1} in our paper, they exclude the possibility of large deviations from Kerr. In the eXTP simulations, the allowed area is only that around $a_* = 0.52$ and $\delta r/r_{\rm Kerr} = 0$. While these simulations include the noise of the instruments, the high flux of XTE~J1752-223 and the large effective area of eXTP lead to $N_{\rm line} \sim 10^7$, which is much higher than $N_{\rm line}$ used in Fig.~\ref{fig1}.

\section{Summary and conclusions}

The aim of this letter was to compare the constraining power from the detection of gravitational waves and the analysis of the iron K$\alpha$ line to test the nature of astrophysical black holes. The gravitational wave case was discussed in Ref.~\cite{kz} and here we have employed the same set-up to study the iron line method. As we have already stressed, both studies should be seen as a preliminary work to provide a rough estimate of the actual capabilities of the two techniques. They both adopt several simplifications that must be removed in future studies.

Our results can at least show that, in the presence of high quality data and with the systematics under control, the iron line method can compete with the gravitational wave approach to test black holes. It is possible that some kinds of deformation can be better constrained by a technique and other kinds of deformations by the other method. Research along these lines is currently underway.

Generally speaking, the observation of a single quasi-normal mode, typically the $l=m=2$ mode, provides two parameters, and we can thus only measure two physical quantities. This is enough to infer the mass $M$ and the spin parameter $a_*$ if we assume the Kerr metric, but it is not if we want to test the nature of a black hole. Even if the accurate measurement of a single quasi-normal frequency could constrain very well some kinds of deformations, it cannot surely do it for others. An accurate measurement of more quasi-normal frequencies may break this parameter degeneracy.

The profile of the iron line has a complicated structure. If properly understood, an accurate measurement of its shape can provide much more details than the measurement of a single quasi-normal mode. However, it is crucial to use the correct astrophysical model. Current models are typically too simple to be considered realistic. For instance, the emissivity profile is usually modelled by two power-law and a breaking radius, which is clearly an approximation. The next generation of X-ray missions, like eXTP, will be able to provide unprecedented high quality data, but it is necessary that at that time we have more sophisticated astrophysical models than the phenomenological ones available today. The weak point of the iron line method with respect to gravitational waves is mainly related to the complication of the astrophysical system. In the case of gravitational waves, the measurement can be relatively clean~\cite{rgw}.


\begin{acknowledgments}
We would like to thank Nicol\'as Yunes for reading a preliminary version of this manuscript and providing useful feedback. We are also grateful to Matteo Guainazzi and James Steiner for valuable comments and suggestions.  The work of C.B. and J.J. was supported by the NSFC (grants 11305038 and U1531117) and the Thousand Young Talents Program. C.B. acknowledges also support from the Alexander von Humboldt Foundation.
\end{acknowledgments}


\end{document}